\documentclass[10pt]{iopart}%
\usepackage{pifont}
\usepackage{graphicx}%
\usepackage{amsfonts}%
\usepackage{amssymb}
\begin{document}

\title{
Magnetic properties of Quantum Corrals from first--principles calculations}
\author{B. Lazarovits\dag 
\footnote[1]{To whom correspondence should be addressed.
E-mail: bl@cms.tuwien.ac.at},
B. \'Ujfalussy\ddag,
L. Szunyogh\dag\S,
B. L.  Gy\"orffy\dag* \ and P. Weinberger\dag}

\begin{abstract}
We present calculations for electronic and magnetic properties of surface
states confined by a circular quantum corral built of magnetic adatoms (Fe) on
a Cu(111) surface. We show the oscillations of charge and magnetization
densities within the corral 
and the possibility of the appearance of spin--polarized states. In
order to classify the peaks in the calculated density of states 
with orbital quantum numbers 
we analyzed the problem in terms of a simple quantum mechanical circular 
well model. 
This model
is also used to estimate the behaviour of the magnetization and energy with
respect to the radius of the circular corral. The calculations are performed
fully relativistically using the embedding technique within the
Korringa-Kohn-Rostoker method.

\end{abstract}

\address{\dag Center for Computational Materials Science,
Technical University Vienna,
A-1060, Gumpendorferstr. 1.a., Vienna, Austria}

\address{\ddag
Metals and Ceramics Division, Oak Ridge National Laboratory,
Oak Ridge, Tennessee 37831, USA }

\address{\S\ Department of Theoretical Physics and Center for Applied
Mathematics and Computational Physics,
Budapest University of Technology and Economics,
Budafoki \'ut 8, H-1111, Budapest, Hungary}

\address{*
H. H. Wills Physics Laboratory, University of Bristol, Bristol BS8 1TL, United Kingdom
}

%\submitto{\JPCM}

\section{Introduction}

Over the past two decades, electrons in two--dimensional (2D) surface states
on closed packed surfaces of noble metals have been at the center of much
experimental and theoretical attention \cite{K83,HP94}. 
For a pristine surface the
energies of such states lay in the 'gap' around the L point of the bulk
Brillouin Zone and the wavefunctions are confined to the surface. The
corresponding dispersion relations have been determined by angle resolved
photoemission spectroscopy and they have been found to be 2D 
free--electron like parabolas \cite{K83,BGO01}. Moreover, they are partially filled and hence
the electrons, which fill them, form a 2D metal. The most
interesting feature of this remarkable state of matter is its response to
perturbations such as caused by placing transition metal atoms on the surface.
As might be expected, such response displays long range, 'Friedel like', charge
oscillations governed by the 2D Fermi 'Surface'. Indeed, one of the iconal
experiments in nano--technology has been the fabrication of a circular
arrangement of 48 Fe atoms on a Cu (111) surface and the direct observation,
by Scanning Tunneling Microscopy (STM), of such oscillations within the
circle \cite{CLE+95,CLE+96,MLE00}. In this paper we wish to discuss the, as yet unexplored,
magnetic properties of such quantum corrals.

Until recently STM studies of atoms on well defined Cu, Ag and Au
(111) surfaces imaged only the charge distribution of the surface
electrons \cite{CLE+95,CLE+96,MLE00}. 
But now, remarkable developments in spin--polarized STM (SPSTM)
\cite{B03}
make the observing of spatial variations in the magnetic density a
distinct possibility and, therefore, an attractive new area of research.
Evidently, this opens up the possibility of building magnetic nanostructures
for both scientific and technological purposes. For instance, while the
observation of a single Fe or Co atom on a Cu surface may be beyond the
spatial resolution of the first generation of SPSTMs, the magnetic state of a
single quantum corral of 50-100 atoms can be readily identified \cite{PKB+04}. 
The motivation behind the
theoretical work reported here is the need to identify the principle
conceptual issues which govern the physics, in general, and the magnetism, in
particular, of such structures.

Individual impurities and clusters of impurities embedded in the above
surface--2D--host--metals have been studied for Friedel oscillations 
around them \cite{PSH+98,DSB+03}, for RKKY interactions between them
\cite{RMM+00,KBE+02,SBT+03} and for a rich variety of Kondo--like
phenomena they are host to \cite{MLE00,PFT01,FHH+01,AS01}. 
Clearly, all these effects can occur inside a corral with
the interesting aspect that now the electronic structure of the host can
be controlled by the geometry of the corral. It is perhaps useful to note that
these circumstances are rather analogous to that of quantum wells in
semiconductor physics \cite{wangbook}.

In the semiclassical limit the states of the quantum corrals can be
associated with classical orbits of particles bouncing off confining walls.
Depending on the shape of the corral the classical motions may be integrable
or chaotic. Thus, quantum corrals can serve as examples of quantum chaos at
work \cite{FH03,SD92}.

Clearly, to think about using different impurities, different
substrates and/or different confining geometries one needs simple but
reliable models as guides. In the unfamiliar physical circumstances at hand, an
efficient way to such models is to perform large scale first--principles
calculations including all conceivably relevant effects and in interpreting the
results in terms of simple models. This is the approach we take in the present
paper. Our first--principles calculations are a spin--polarized and relativistic
generalization of the pioneering work of H\"ormandinger and Pendry 
\cite{HP94} and
Crampin and collaborators \cite{CB96,papers}. We interpret our 
results in terms of a flat bottom 'circular potential well',
non--relativistic, model. 
To illustrate the power of this
approach, once it has been established that the model faithfully reproduces
the main features of the results from first--principles calculations,
properties which would be too difficult 
to calculate from first principles are estimated.

\section{Method of calculations}
\label{sec:theo}

Within multiple scattering theory of the electronic structure the
information about each atom (scattering center) is coded in the
scattering path operator (SPO) matrix,
$\mbox{\boldmath $\tau$}(E)=\{\underline{\tau}^{nm}(E)\}=\{\tau_{QQ^{\prime}%
}^{nm}(E)\}$, with $Q$ and $Q^{\prime}$ being angular momentum indices
referring to atomic sites whose position vectors are
labeled by $n$ and $m$, respectively, and $E$ being the energy. For
scatterers described by non--overlapping (muffin--tin) potential wells,
the SPO matrix,
\begin{equation}
\mbox{\boldmath $\tau$}(E)=[\mbox{\boldmath $t$}^{-1}(E)-\mbox{\boldmath
$G$}(E)]^{-1}\quad,\label{eq:realtau}%
\end{equation}
describes the full hierarchy of scattering events between any two
particular sites, $n$ and $m$. In \Eref{eq:realtau}, 
$\mbox{\boldmath
$t$}(E)=\{\underline{t}^{n}(E)\,\delta_{nm}\}=\{t_{QQ^{\prime}}^{n}%
(E)\,\delta_{nm}\}$ and $\mbox{\boldmath $G$}(E)=\{\underline{G}%
^{nm}(E)\}=\{G_{QQ^{\prime}}^{nm}(E)\}$ denote the single--site \emph{t}%
-matrices on the energy shell and the real-space structure 
constants, respectively \cite{weinbook}. 

We begin our investigations by two fully self--consistent calculations: one
for a semi--infinite Cu with a (111) surface and another one for a single Fe
adatom on this semi--infinite host. In the first case, above the
Cu layers there are two layers of sites
occupied by atomic cells without nuclear charge which
we call empty sites, but we note that they do contain electronic
charge and their electrostatic potential is calculated fully
self--consistently. In the second type of self--consistent calculation, 
one empty site above the topmost
Cu layer is occupied by an Fe atom.
We then construct 'crystal potentials' for further single
pass, one-electron calculations according to the following recipe: 
within the corresponding atomic spheres, 
all Fe sites are described by the potential
as obtained from the single impurity calculation and all other sites are
described by the appropriate potentials from the pristine surface
calculation. The configurations of interest are those in which 
some empty sites in the layer above the topmost Cu layer are replaced
by Fe atoms, forming thus the wall of a corral
as shown in \Fref{fig:geo}. 
Clearly, by this construction, at the level of
the 'crystal potential', we are neglecting the influence of neighbouring Fe
atoms on each other. Fortunately, 
this 'frozen potential' approximation was shown to
be reasonable in the case of Fe adatoms on Ag(100) \cite{LSW02}.

For a given 'crystal potential' constructed according to the above recipe 
we solve the multiple scattering
problem by means of the embedding method \cite{LSW02}. 
In short, we embed a cluster of sites labeled by
$\mathcal{C}$, consisting of the corral Fe atoms and selected empty
sites inside and outside the corral, in the
unperturbed semi--infinite Cu host. A particular cluster $\mathcal{C}$ can then
be treated as perturbation of the host. In practice, we first calculate the
SPO matrix of the 2D translational invariant layered host, 
${\mbox{\boldmath $\tau$}}%
_{h}(\mathbf{k}_{\parallel},E)=\{\underline{\tau}_{h}^{pq}(\mathbf{k}%
_{\parallel},E)\}$, within the framework of the SKKR method \cite{SUW95},
where $p$ and $q$ denote layers and the $\mathbf{k}_{\parallel}$ are vectors
in the surface Brillouin zone (SBZ). The real--space SPO matrix is then given by
\begin{equation}
\underline{\tau}_{h}^{mn}(E)=\frac{1}{\Omega_{SBZ}}\int\limits_{SBZ}%
e^{-i(\mathbf{T}_{i}-\mathbf{T}_{j})\mathbf{k}_{\parallel}}\underline{\tau
}_{h}^{pq}(\mathbf{k}_{\parallel},E)d^{2}k_{\parallel}\quad,\label{eq:bzint}%
\end{equation}
where the atomic position vectors are decomposed as $\mathbf{R}_{m}%
=\mathbf{T}_{i}+\mathbf{c}_{p}$ and $\mathbf{R}_{n}=\mathbf{T}_{j}%
+\mathbf{c}_{q}$ with $\mathbf{T}_{i}$ and $\mathbf{T}_{j}$ being 2D lattice
vectors, $\mathbf{c}_{p}$ and $\mathbf{c}_{q}$ the so--called
layer--generating vectors, and $\Omega_{SBZ}$ is the unit area of the surface
Brillouin zone.

By replacing the \emph{t}--matrices of the unperturbed host,
$\mbox{\boldmath
$t$}_{h}(E)$, with those of the cluster-atoms,
$\mbox{\boldmath $t$}_{\mathcal{C}}(E)$, leads to the following Dyson like
equation,
\begin{equation}
\mbox{\boldmath $\tau$}_{\mathcal{C}}(E)= \mbox{\boldmath $\tau$}_{h}(E)
\left[  \mbox{\boldmath $I$}- (\mbox{\boldmath $t$}^{-1}_{h}(E) -
\mbox{\boldmath $t$}^{-1}_{\mathcal{C}}(E)) \mbox{\boldmath $\tau$}_{h}(E)
\right] ^{-1} \quad,\label{eq:dyson}%
\end{equation}
where $\mbox{\boldmath $\tau$}_{\mathcal{C}}(E)$ is the SPO matrix
corresponding to all sites in cluster $\mathcal{C}$, from which in turn 
local quantities, such as the densities of states (DOS), 
magnetic densities of states (MDOS), 
spin and orbital moments, as well as the total energy can be calculated.
Note, that \Eref{eq:dyson} takes into account all scattering events both
inside and outside the cluster.

\subsection{Geometries of interest}

\label{sec:geo}

\begin{figure}[ptbh]
\begin{center}
\includegraphics[width=0.80\textwidth,clip]{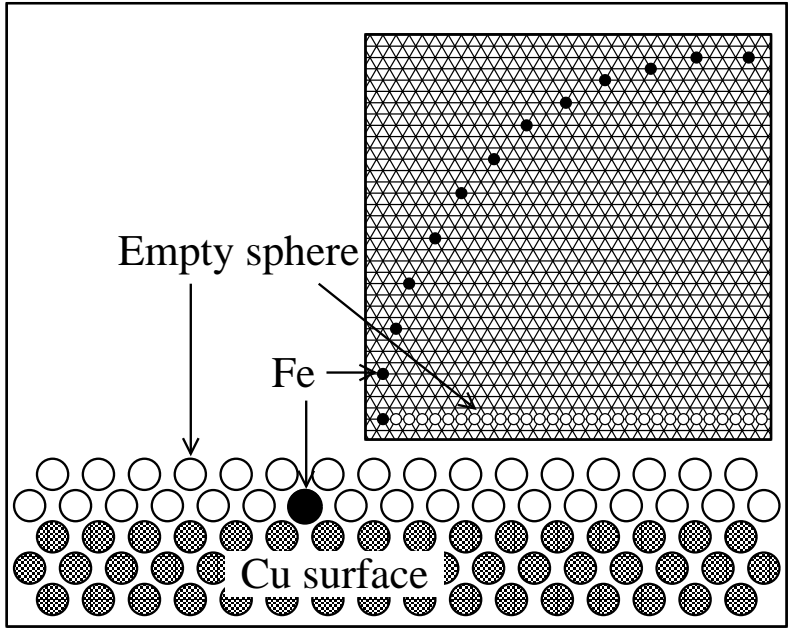}
\end{center}
\par
\vskip -10pt
\caption{Cross section of the surface showing the position of sites
in the 'vacuum layers' (open circles), the Cu surface layers 
(gray circles)
and an Fe impurity in the first 'vacuum layer' (black circle). 
Inset: the positions of the Fe atoms (black 
dots) and the empty spheres (open circles) along a diameter for a 
quadrant of the investigated
corral. }%
\label{fig:geo}%
\end{figure}

The 48 Fe atoms forming the corral were positioned on the surface along a
circle with a diameter of 28~$a$, where $a$ is the 2D lattice
constant of the fcc(111) Cu surface. The investigated geometry is shown in
the inset of \Fref{fig:geo}. This is similar to
a popular experimental one \cite{CLE+95}. The corral sites refer to the positions of
an ideal fcc parent lattice with the experimental Cu lattice constant,
therefore, there is some deviation from the exact circular shape. 
The positions
of the adatoms were chosen such that the spacing between them is
approximately constant. Within the interior of the corral
the physical properties (DOS, MDOS) of 55 empty spheres along a diameter were
calculated. Note that the rotational symmetry of the considered
structure is not continuous. Nevertheless, to reduce the computational
effort, we followed Crampin \emph{et~al.}~\cite{CB96} and assumed that the
properties on an arbitrary position within the corral depend only on the
distance from the center of the circle.

\subsection{Computational details}

\label{sec:comdet} Self--consistent, fully relativistic calculations 
for the pristine Cu(111) surface as well as for the
Fe adatom on Cu(111) have been performed in the framework of the local
spin--density approximation (LSDA) as parameterized by Vosko \emph{et al.}
\cite{VWN80}.  The potentials were treated within the atomic sphere 
approximation (ASA). For the
calculation of the $t$--matrices and for the multipole expansion of the charge
densities, necessary to evaluate the Madelung potentials, a cut--off of
$\ell_{max}=2$ was used. The energy integrations were performed by sampling
16 points on a semicircular contour in the complex energy plane 
according to an asymmetric Gaussian quadrature. Both for the self--consistent
calculation of the Cu(111) surface and for the evaluation of
\Eref{eq:bzint} we used 70 $k_{\parallel}$--points in the irreducible
wedge of the SBZ (ISBZ). The DOS and MDOS were calculated at an energy mesh parallel
to the real axis with an imaginary part of 0.5~mRyd. In here, in order to cover the
surface--state properties properly, a sampling of about 3300 ${k}_{\parallel}$--points
within the ISBZ was necessary.

\section{Results of first-principles calculations}

\label{sec:res}

\subsection{Clean Cu(111) surface}

\begin{figure}[ptbh]
\begin{center}
\includegraphics[width=0.80\textwidth,clip]{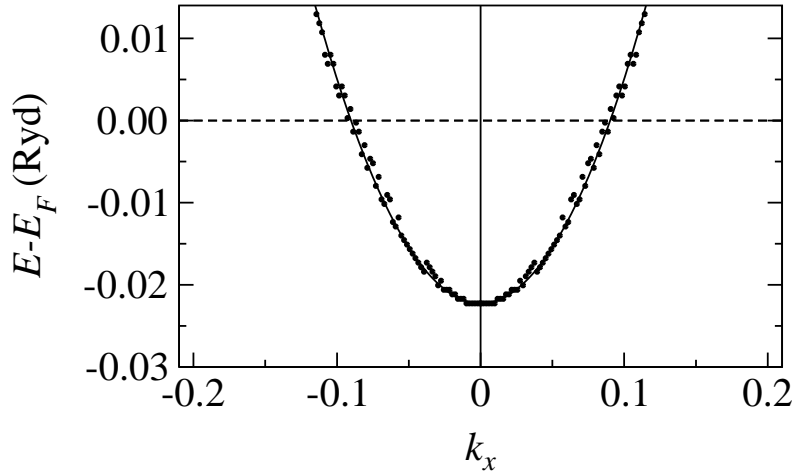}
\end{center}
\par
\vskip -10pt
\caption{Dots: Bloch spectral function maxima near to the
$\bar{\Gamma}$ point of the SBZ. Line: parabola fitted to the calculated
maxima. It should be noted that only the first third of the SBZ is 
displayed ($\bar{K}\approx0.65/a.u.$). The
estimated effective mass is $m^{\ast}=0.366 \,m_e$. }%
\label{fig:band}%
\end{figure}

In order to determine the dispersion relation and the effective mass of the
surface electrons the \textit{Bloch-spectral function} (BSF) \cite{weinbook}
was calculated
between $\bar{\Gamma}$ and $\bar{K}$ in the fcc(111) SBZ close to the Fermi
energy by using the SKKR method, in which the properties of the semi--infinite
substrate are calculated by the surface Green function method \cite{SUW95}.
The proper treatment of the host is necessary to account for
the interaction between the bulk and surface states in an ab--initio way. The
maxima of the BSF can be identified as the surface state band. In agreement
with the experiments, the calculated dispersion relation is free-electron like
and can be estimated with a parabola as indicated in \Fref{fig:band}. The
bottom of the calculated surface states band, $E_{B}$, is 0.3~eV
below the Fermi energy which is a bit smaller than the experimental value
(0.39~eV) \cite{K83,BGO01}. By using a quadratic approximation for the
dispersion relation,
\begin{equation}
E(\mathbf{k}_\parallel)=E_{B}+\frac{\hbar^{2}k_\parallel^{2}}{2m^{\ast}}+\dots
\quad,\label{eq:Est}%
\end{equation}
we obtained an effective mass with $m^{\ast}=0.366 \,m_e$ which is in
good agreement with the experiment
($m^{\ast}/m_e \approx 0.41$) \cite{BGO01}. 

\subsection{\bigskip Fe impurity on a Cu(111) surface}

In these calculations the magnetic moment of the Fe impurity on the surface
turned out to be 3.27 $\mu_{B}$ and, due to the magneto--crystalline
anisotropy (MCA) induced by spin--orbit coupling \cite{Bruno93}, 
its preferred orientation was perpendicular to the surface.
The MCA energy, defined as the difference of the LSDA total energy between an
in--plane and a normal--to--plane orientation of the magnetization, 
yielded 4.3 meV.
Both of these results are consistent with those already reported in 
the literature \cite{LSW+03}.
We note that the orientation of the magnetic moments in the quantum corral
built up from Fe atoms is, in principle, 
also affected by the so--called shape anisotropy,
arising from the magnetostatic dipole--dipole energy \cite{Bruno93}. 
This, purely classical, interaction would 
direct the orientations of the magnetic moments into
the plane.  According to
our estimates, the magnetostatic dipole--dipole energy 
is, however, at least by one order less in magnitude 
than the above MCA energy.
Therefore, in our further calculations the Fe adatoms 
were taken to be 
spin--polarized in the $z$ direction, i.e., normal to the surface.
Moreover, the exchange (RKKY) interaction between two Fe atoms can be both
ferro-- and antiferromagnetic depending on the distance between them.
This implies that by varying the geometry of the corral various ground--state 
magnetic configurations can occur. Because of experimental interest,
in this work we have studied the case of a ferromagnetic corral. 

\vspace{0.5cm}
\begin{figure}[pthb]
\begin{center}
\includegraphics[width=0.50\textwidth,clip]{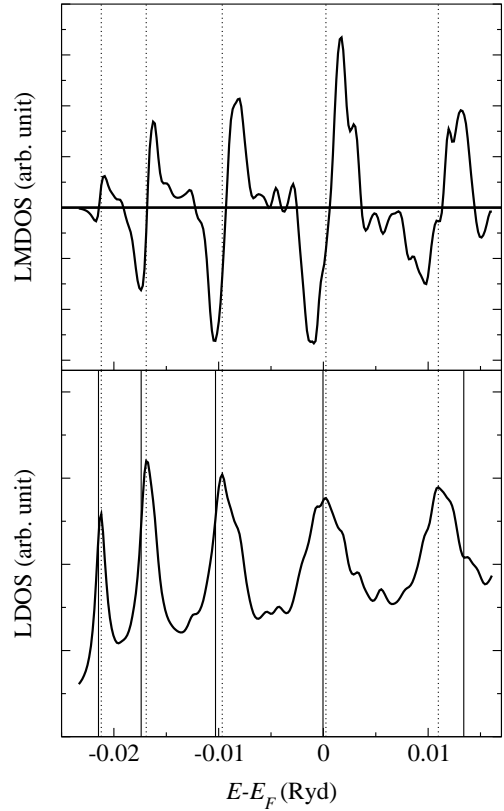}
\end{center}
\par
\vskip -10 pt\caption{Calculated local density of states (LDOS) and 
local magnetic density
of states (LMDOS) at the center of a quantum corral. The dotted lines indicate
the maxima in the LDOS. Vertical solid lines are the energy eigenvalues,
$E_{0i}$, predicted by the circular quantum well model (see \Sref{boxmodel}). }%
\label{fig:dos_mid}%
\end{figure}

\begin{figure}[ptbh]
\begin{center}
\includegraphics[width=0.50\textwidth,clip]{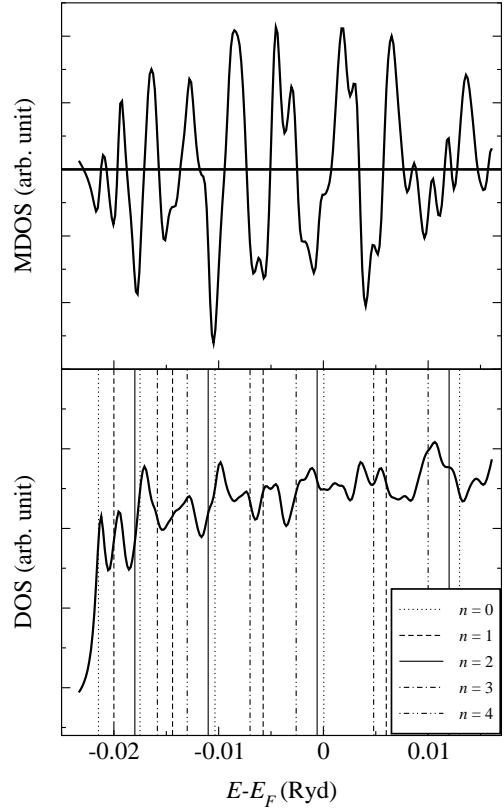}
\end{center}
\par
\vskip -10 pt\caption{Calculated density of states (DOS) and magnetic density
of states (MDOS) summed for a diameter of the corral. Vertical lines
indicate the energy eigenvalues 
corresponding to different values of the quantumnumber $n$ 
predicted by the circular quantum well model 
(see \Sref{boxmodel}). }
\label{fig:sumdos}%
\end{figure}

\begin{figure}[ptbh]
\begin{center}
\includegraphics[width=0.70\textwidth,clip]{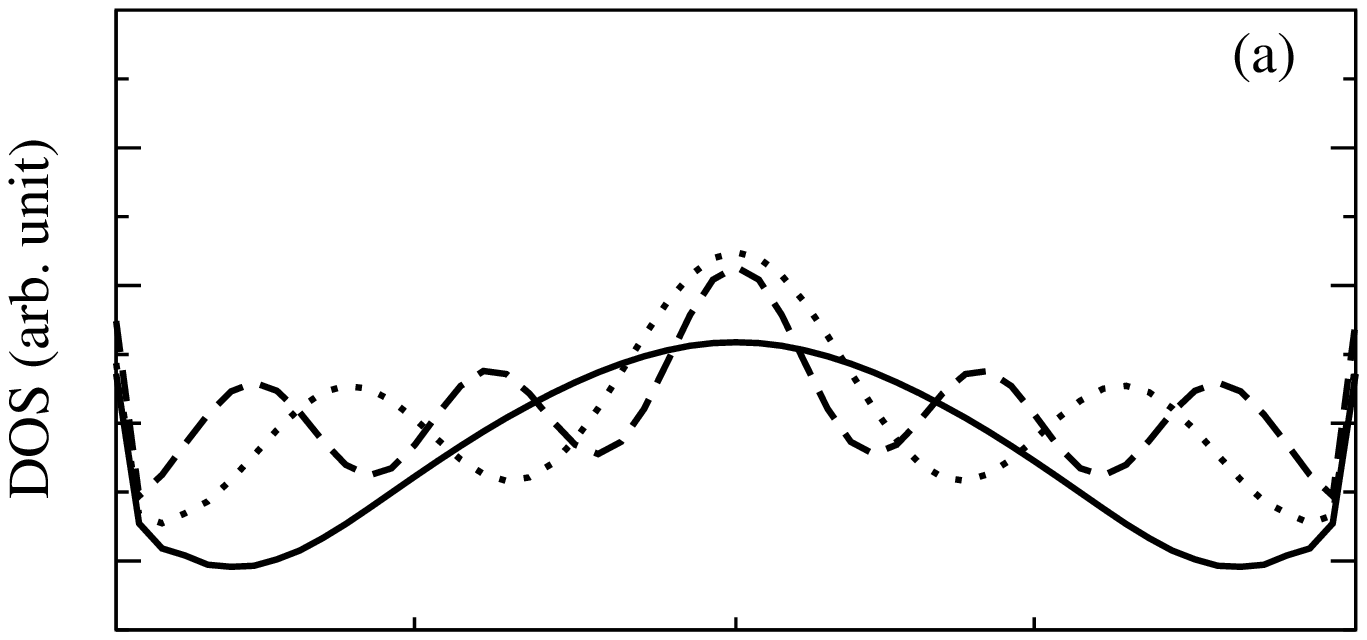}
\includegraphics[width=0.70\textwidth,clip]{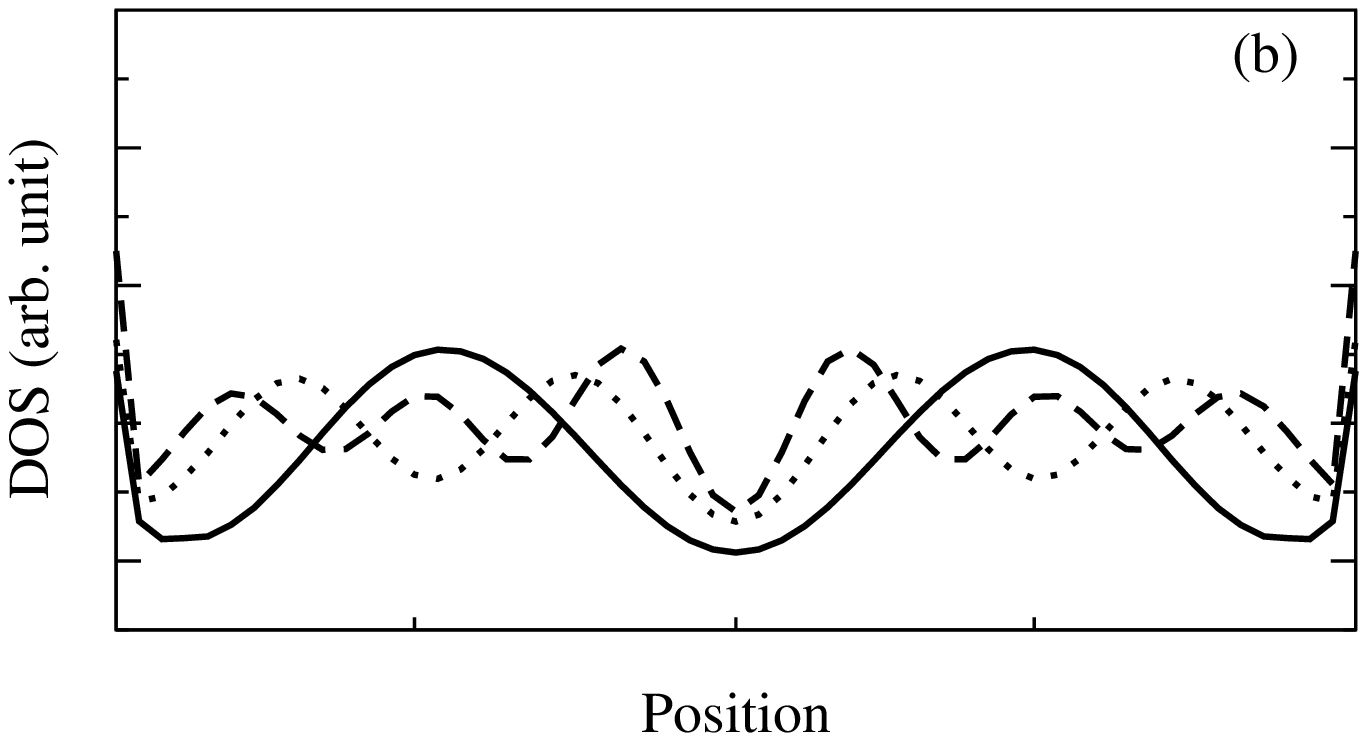}
\end{center}
\par
\vskip -10pt\caption{Spatial distribution of the DOS along
a diameter of the quantum corral at energies 
corresponding to selected peaks in \Fref{fig:sumdos}. 
The line shapes in part (a) and (b) can be identified as $n=0$ 
and as $n\neq0$ quantumstates of a circular quantum well model
(see \Sref{boxmodel}), respectively.}
\label{fig:spatdos}%
\end{figure}

\begin{figure}[ptbh]
\begin{center}
\includegraphics[width=0.80\textwidth,clip]{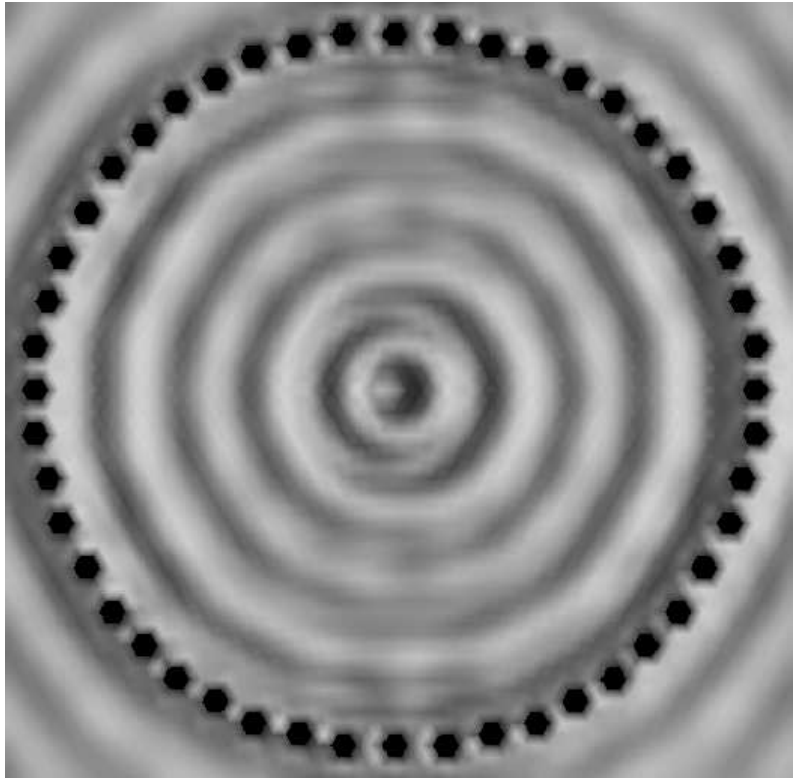}
\end{center}
\par
\vskip -10pt\caption{Spatial distribution of the DOS at the energy 
corresponding to
the fifth peak of the LDOS at the central position ($E-E_F \simeq$ 0.01 Ryd).
The LDOS of the Fe atoms is removed from the figure.
}%
\label{fig:spatdos2}%
\end{figure}

\subsection{DOS of confined surface states}

In order to study the properties of surface states confined by the quantum
corral first we investigated the local density of states (LDOS),
\begin{equation}
n_{i}(E)=n_{i,\uparrow}(E)+n_{i,\downarrow}(E)=-\frac{1}{\pi}\sum
_{\sigma=\uparrow,\downarrow}\mathrm{Im}\mathrm{Tr} \: G_{\sigma\sigma}^{ii}(E)
\quad ,
\end{equation}
and the local magnetic density of states (LMDOS),
\begin{equation}
m_{i}(E)=n_{i,\uparrow}(E)-n_{i,\downarrow}(E)
\quad ,
\end{equation}
of an \textit{empty sphere} at various lattice points (sites)
labeled by $i$ within the
corral, where $G_{\sigma\sigma}^{ii}(E)$ is the site-- and spin--diagonal 
part of the resolvent in $(\ell,m,\sigma)$ representation. Although in our
relativistic theory the spin is not a good quantumnumber, and hence
$G_{\sigma\sigma^{\prime}}^{ii}(E)$ is not diagonal, 
we can interpret the diagonal elements in a similar manner 
as in a non--relativistic
theory, because within an empty sphere the spin--orbit coupling is bound to
be small. In \Fref{fig:dos_mid} we show the LDOS and LMDOS at the
center of the corral.

The striking peaky structure of the DOS in \Fref{fig:dos_mid} 
is in sharp contrast to the
constant 2D density of states expected from the dispersion relation in
\Eref{eq:Est}. However, the peaks are rather similar to those found by
Crampin \emph{et~al.} \cite{CB96} who 
interpreted them as 'bound states' within the corral. 
We confirm this interpretation using a simple circular well model in
\Sref{interpretation}.
Nevertheless, it is somewhat surprising that
the coherent scattering from a circular arrangement of Fe impurities is
almost equivalent to that of an infinite confining potential wall. 
Evidently, such confined states are analogous to the quantum well states 
in semiconductor physics \cite{wangbook}. 
In contrast to Crampin \emph{et~al.} \cite{CB96}, we have also 
calculated the spin--resolved densities of states (LMDOS) 
which are also presented in \Fref{fig:dos_mid}. 
Clearly, the LMDOS is more structured than the LDOS suggesting that 
the quantum well states are exchange split. 
To lend further credence to such an interpretation we calculated
the sum of the LDOS and LMDOS,
\begin{equation}
n(E)=\sum_{i}n_{i}(E) \quad ,\quad m(E)=\sum_{i}m_{i}(E) \quad,
\end{equation}
 along a diameter of the corral. 
Note that near to the Fe atoms the LDOS increases rapidly
due to the direct charge transfer, therefore, in the
above sums the contributions from the empty spheres adjacent to the Fe atoms
are neglected. 
These summed densities of states are then plotted in \Fref{fig:sumdos}. 
Although the peaks seen in \Fref{fig:dos_mid} are still present,
\Fref{fig:sumdos} suggests a more complex spectrum. In order to shed light 
on the nature of the extra states, the spatial resolution of the DOS
at selected energies corresponding to the most prominent peaks are shown in
\Fref{fig:spatdos} and, for the fifth peak in \Fref{fig:dos_mid} 
($E-E_F \simeq$ 0.01 Ryd), the LDOS for the whole area 
within the corral is depicted in \Fref{fig:spatdos2}. 
Note that the oscillations continue outside the corral. This
implies that the states, we have been studying, are really resonances
rather than bound states.
Reassuringly, the obtained pattern for the confined surface 
electrons agrees well with the experimental one \cite{CLE+95,CLE+96}.
Beyond agreeing with experiments,
the oscillations with distance from the center in 
\Fref{fig:spatdos} strongly support the interpretation that the 
selected peak positions are those of 'bound' states confined by the corral.

Evidently, even for the present very simple geometry the full results are
too complicated to allow an unambiguous identification of each 
structure with specific
physical processes. To maximize the information gained from our
first--principles calculations we shall now introduce a simple model whose
parameters can be chosen such that it reproduces most of the above results
quantitatively.

\subsection{The non--relativistic Circular Quantum Well model}
\label{boxmodel}

\begin{figure}[htbp]
\begin{center}
\includegraphics[width=0.50\textwidth,clip]{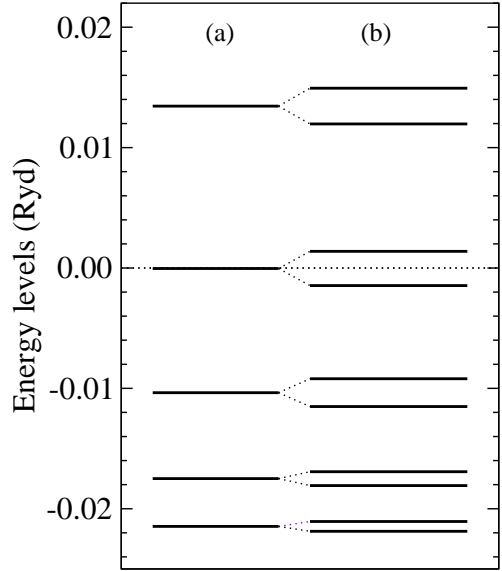}
\end{center}
\vskip -10pt
\caption{Energy levels of the \(n=0\) states within
the circular quantum well model at the radius 
used for the self-consistent calculation. (a): non--spinpolarized
model, (b): spin--splitting term added.
The energy dependent splitting between the spin--up and spin--down states
is estimated from the LMDOS shown in \Fref{fig:dos_mid}.
}
\label{fig:levels}
\end{figure}

The circular quantum well model is defined by a potential of the form,
\begin{equation}
V(\mathbf{r})=\left\{
\begin{array}
[c]{ccc}%
0 & \mathrm{if} & r<r\\
+\infty & \mathrm{if} & r\geq R
\end{array}
\right. \quad .
\end{equation}
It should be recalled that the radial solutions of the 
corresponding Schr\"odinger equation are
Bessel function of the first kind $J_{n}(pr)$, where
\begin{equation}
p=\sqrt{\frac{2m^{\ast}}{\hbar^{2}}E}\quad.
\end{equation}
The energy eigenvalues arise from the boundary condition that the radial
solutions vanish at the boundary:
\begin{equation} \label{eq:eni-nonspinpol}
E_{ni}=\frac{x_{ni}^{2}}{R^{2}}\frac{\hbar^{2}}{2m^{\ast}}\quad,
\end{equation}
where $x_{ni}$ refer to the $i$th zero of the Bessel function $J_{n}$. 
In order to
investigate the magnetic properties of this model, one can generalize it 
by adding
a spin--dependent part, $V_{\uparrow(\downarrow)}(r)=V(r)+U_{i,\uparrow
(\downarrow)}$, where the constant 
$U_{i,\uparrow}=-U_{i,\downarrow}$ can also depend
on the quantumnumber \(n\). 
The magnetic exchange term modifies the energy values as follows,
\begin{equation} \label{eq:eni-spinpol}
E_{ni}^{\uparrow(\downarrow)}=\frac{x_{ni}^{2}}{R^{2}}\frac{\hbar^{2}%
}{2m^{\ast}}+U_{i,\uparrow(\downarrow)}\quad.
\end{equation}
In order to facilitate a comparison with the ab--initio results,
the value of $R=27$~$a$ is used. 
This means that the radial solution has to vanish at the
lattice position neighbouring the corral atoms. The corresponding energy
values with and without magnetic exchange term for $n=0$ can be seen in
\Fref{fig:levels}.

\vspace{0.5cm}

\subsection{Interpretation of the results of first--principles calculations}
\label{interpretation}

In what follows we comment on the results of our first--principles
calculations in the light of the above circular quantum well model. 
Firstly, we note that
in \Fref{fig:dos_mid} the vertical solid lines correspond to the bound state 
energies $E_{0i}$ for $m^{\ast}/m_{e}=0.366$ and 
$U_{i,\uparrow(\downarrow)}=0$. 
The agreement between the peak positions and the $E_{0i}$'s
is especially striking in the low-energy regime. This can be viewed
as an indication that at lower energies the scatterers act more like a
hard wall than at higher energies.
Given that the model is solved by using the Schr\"odinger equation,
it also suggests that relativistic effects inside the corral are not
too important.
In the first--principles calculations the peaks have finite widths
due to the combined effects of the discreteness of the confining boundary and
the energy dependent scattering into bulk states. Remarkably, the
width of the peaks agrees quantitatively with the experimental values
\cite{CLE93b}.
This result, however, may be fortuitous. Note, for instance, that
for reasons not entirely clear, in the experiments the third
peak is the highest one, but in our result the highest peak is the second one.

The spatial distribution of the DOS along a diameter of the corral  
is plotted in the upper panel of
\Fref{fig:spatdos} at energies referring to the first three peaks 
in \Fref{fig:dos_mid}.
As can be expected the LDOS has a maximum at the
center similarly to the $J_{n=0}$ Bessel function. The circular well
model suggests that there are states with non--zero orbital moment, $n\neq0$,
for which there is a minimum at the center of the corral. In order to
find these states, in \Fref{fig:sumdos} we investigated 
the spatial sum of the LDOS along a diameter.
Evidently, the $n\neq0$ states would contribute to this sum.
Indeed, as noticed already, in \Fref{fig:sumdos} there
are new peaks as compared with \Fref{fig:dos_mid}
to which states with zero amplitude at the origing do not contribute.
In \Fref{fig:sumdos},
the values $E_{ni}$ as predicted from the circular quantum well model
with zero exchange--split term, $U_{i\uparrow (\downarrow)}=0$,
are also indicated by vertical lines up to $n=4$. Note, for example, that
the second peak clearly corresponds to the $n=1, i=1$ quantum state.
To pursue this matter further, in the lower
panel of \Fref{fig:spatdos}, we plotted the spatial density of the
states corresponding to those peaks which appear only in the summed DOS.
Reassuringly, these spatial densities have minimum at the center.
For higher energies this correspondence is not so clear, nevertheless, 
this comparison serves as an explanation why the 
shape of the peaks differs from that of an ideal Lorentzian. 
The small features between
the prominent peaks which can be seen in \Fref{fig:dos_mid}, and in the
experimental results \cite{CLE+95} also show some systematic trends which
we assume can be explained n terms of a the circular quantum well model
based on a two--dimensional Dirac equation.

Turning to the LMDOS at the central position, we note that 
in the upper panel in \Fref{fig:dos_mid}
the prominent DOS peaks are split into a pair of 'up' and 'down' peaks. 
This oscillatory behaviour of
the LMDOS is a consequence of the spin--polarization of the corral Fe atoms,
namely of the perturbation of the surface-state electrons by the Fe atoms
forming the corral. This exchange splitting can be
reproduced in the circular well model by choosing 
$U_{i,\uparrow(\downarrow)}$ appropriately. 
It turns out that for a good fit
$U_{i,\uparrow(\downarrow)}$ should be non--uniform in energy. 
The corresponding energy values with the exchange splitting terms
estimated from \Fref{fig:dos_mid} are shown in \Fref{fig:levels}%
b. Thus, the prediction of the model is that if the Fermi energy falls
between an exchange--split doublet then the whole corral has a net
magnetic moment of one Bohr magneton in addition to the magnetization due
to the Fe atoms forming the wall. 
Furthermore, such a moment will not be
uniformly distributed within the corral but varies from empty--cell to
empty--cell as required by the wavefunction of the confined surface
state, namely,
it oscillates like the charge density in \Fref{fig:spatdos}.
Of course,
even for a relatively small corral of $R=27$~$a$ , as indicated by the 
highly structured summed LMDOS in the upper panel of \Fref{fig:sumdos}, 
there are many
states near the Fermi energy and hence the local magnetic moment can vary
rapidly with Fermi energy and spatial location.
Some of the complexities in
this figure can be attributed to the appearance of the $n\neq0$
states. Although the exchange splitting around the DOS peaks can be
clearly observed, a full analysis of the results of our relativistic 
spin--polarized ab--initio calculations will have to has to be 
based on a relativistic treatment of the circular well model.

\vspace{0.5cm}
\begin{figure}[ptbh]
\begin{center}
\includegraphics[width=0.60\textwidth,clip]{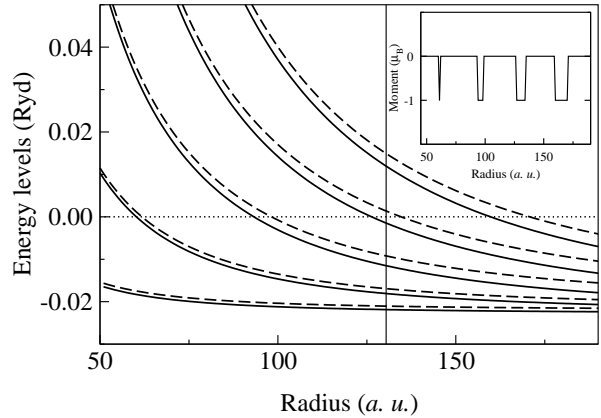}
\end{center}
\par
\vskip -10pt\caption{The $n=0$ energy levels with spin--splitting (dashed line:
spin--up states, solid lines: spin--down states). Inset: Value of the
magnetization with respect to the radius within the circular
quantum well model due to the $n=0$
states. The energy dependent splitting between the spin--up and spin--down
states is estimated from the MDOS calculations. The vertical line indicates
the radius used for the first--principles calculation. }%
\label{fig:levelsr}%
\end{figure}

\subsection{Tuning of the magnetic properties}

From the point of view of engineering corrals with specific properties it is
important to investigate the magnetic properties of the confined surface
states within the corral as a function of the corral radius $R$.
As a preliminary effort in this direction we shall now make some
estimates based on the model whose credibility we have established in the
previous sections. Firstly, we have calculated the dependence of the
exchange split energy levels, shown in \Fref{fig:levels} for the
specific case of $R=130.41$~$a.u.$, as a function of $R$. The
results are displayed in \Fref{fig:levelsr}. 
Assuming that the exchange
energies are independent from the geometry we used the values estimated from
\Fref{fig:dos_mid}. 
It can be seen that as the radius is
decreased the energy levels are pushed upwards, possibly
changing the number of the occupied states. 
As a
consequence of this effect one can find ranges of radii where the spin--down
state is occupied but the corresponding spin--up state is empty and, 
therefore, the surface states hold a finite magnetic moment.
The predicted total magnetic moment of the surface states is 
depicted in the inset of \Fref{fig:levelsr}. 
It should be stressed that the model neglects
a multitude of effects such as the partial confinement of the electrons and the
width of the levels. In realistic first--principles calculations the
magnetic moment is expected to show a smoother change with the radius. 
When states with $n>0$ are also taken into account the results 
are expected to be even more complicated but, qualitatively, 
the basic effect should remain the same. For instance, at the center
of the corral one can expect from \Fref{fig:sumdos} that it is enough to
take into account the $n=0$ states. Therefore, we can predict
that at least at the center there is a finite magnetic moment at well
defined geometries. In short, by varying the geometry a rich variety of
magnetic states can be produced.

\begin{figure}[ptbh]
\begin{center}
\includegraphics[width=0.60\textwidth,clip]{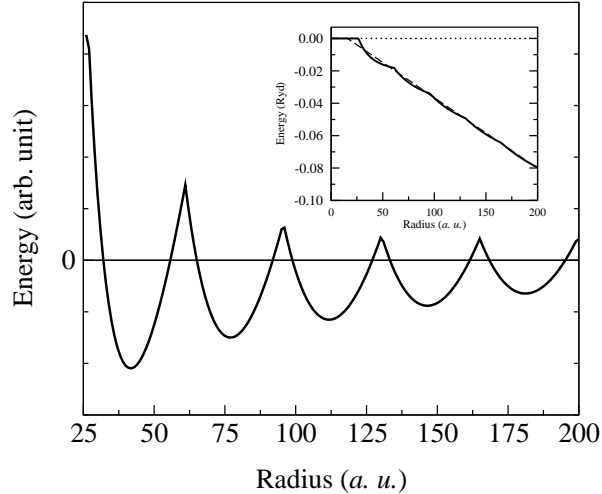}
\end{center}
\par
\vskip -10pt\caption{Oscillation in the total energy with respect to the radius
of the corral within the circular quantum well model due to the $n=0$ states. 
In the inset, the total
energy and its best linear fit \((R>25 \, a.u.)\) 
are plotted by solid and dashed line, respectively.
In the main figure, the difference between the total energy and its 
linear fit
are depicted for a better representation of the oscillations. }%
\label{fig:energosc}%
\end{figure}

As might be expected by now the total energy also shows an interesting
behaviour by varying the corral radius. As a simple, and perhaps artificial,
example we have studied the case when symmetry constrains the system
such that only the $n=0$ states are occupied. 
The total energy for this case is shown in \Fref{fig:energosc}. 
The oscillations resemble and have similar
origin as those responsible for the de~Haas--van~Alphen effect.
Note, however, that in the present example the oscillations are not 
equally spaced but follow from the distribution of the Bessel zeros.
Interestingly, similar oscillations can be expected if, instead the 
radius $R$, the Fermi energy $E_F$ changes. In an experiment one may contrive 
such variations in $E_F$ by 'gating' the corral with an STM tip.

\section{Summary}

In this work we presented calculations of the electronic and magnetic
properties of the surface states confined in a circular quantum corral. The
ab--initio results are interpreted in terms of a simple quantum
mechanical, circular potential well model with infinitely high walls.
We found that at low energies the energy levels of the model gave a good
quantitative account of the peaks of the DOS obtained in 
ab--initio calculations.
In particular, unlike previous calculations for the quantum corrals, we were
able study and interpret the magnetic as well as the charge oscillations
within the corral.
On the bases of these calculations we conclude that a rich
variety of magnetic structures can be expected by varying the shape, size
and gating of these fascinating nanostructures.

\ack Financial support was provided by the Center for Computational Materials
Science (Contract No. GZ 45.531), the Austrian Science Foundation (Contract
No. W004), the Research and Technological Cooperation Project between Austria
and Hungary (Contract No. A-3/03) and the Hungarian National Scientific
Research Foundation (OTKA T046267 and OTKA T037856). The work of BU 
was supported by DOE-OS, BES-DMSE under contract number DE-AC05-00OR22725 with
UT-Battelle LLC.

\end{document}